 \documentclass[nohyper,12pt,letterpaper]{JHEP3}
 \usepackage{graphicx}
 \usepackage{amsmath}
 
 \def\bR{{\mathbb R}}
 \def\bP{{\mathbb P}}

 \def\bZ{{\mathbb Z}}

 \newcommand{\vev}[1]{{\left< {#1} \right>}}

\title{Exact probes of orientifolds}

\author{Bartomeu Fiol, Blai Garolera and Gen\'is Torrents  \\

Departament de F{\'\i}sica Fonamental i \\Institut de Ci{\`e}ncies del Cosmos, 

Universitat de Barcelona,

Mart{\'\i}\ i Franqu{\`e}s 1, 08028 Barcelona, Catalonia, Spain \\

\email{bfiol@ub.edu, bgarolera@ffn.ub.es, genistv@icc.ub.edu }}

\abstract{We compute the exact vacuum expectation value of circular Wilson loops for Euclidean ${\cal N}=4$ super Yang-Mills with $G=SO(N),Sp(N)$, in the fundamental and spinor representations. These field theories are dual to type IIB string theory compactified on $AdS_5\times {\mathbb R} {\mathbb P}^5$ plus certain choices of discrete torsion, and we use our results to probe this holographic duality. We first revisit the LLM-type geometries having $AdS_5\times {\mathbb R} {\mathbb P}^5$ as ground state. Our results clarify and refine the identification of these LLM-type geometries as bubbling geometries arising from fermions on a half harmonic oscillator. We furthermore identify the presence of discrete torsion with the one-fermion Wigner distribution becoming negative at the origin of phase space. We then turn to the string world-sheet interpretation of our results and argue that for the quantities considered they imply two features: first,  the contribution coming from world-sheets with a single crosscap is closely related to the contribution coming from orientable world-sheets, and second, world-sheets with two crosscaps don't contribute to these quantities.}  

\begin{document}
\section{Introduction}
The AdS/CFT correspondence has drastically changed our view on the interrelations between field theory and quantum gravity. However, at the level of specific results, it seems fair to assess that it has not brought as many new results in quantum gravity as in field theory. Indeed, while it has allowed access to regimes of field theory previously unexplored, the amount of work using field theory results to learn about quantum gravity has been smaller. One of the main reasons of this state of affairs is of course the paucity of known results in the relevant regimes of field theory. 

Localization has emerged as a powerful technique to drastically simplify very specific computations in supersymmetric field theories, allowing in some cases to obtain exact results \cite{Pestun:2007rz, Kapustin:2009kz, Drukker:2010nc, Jafferis:2010un}. In particular, for 4d ${\cal N}=2$ super Yang Mills theories with a Lagrangian description, the evaluation of the vev of certain circular Wilson loops boils down to a matrix model computation \cite{Pestun:2007rz}. Furthermore, for the particular case of ${\cal N}=4$ SYM, the matrix model is Gaussian \cite{Erickson:2000af,Drukker:2000rr, Pestun:2007rz}, so all the integrals can be computed exactly. This has been done for $G=U(N),SU(N)$ first for a Wilson loop in the fundamental representation, and more recently for other representations \cite{Gomis:2009ir, Fiol:2013hna}. Even  though the quantities that can be computed thanks to localization must satisfy a number of conditions that make them non-generic,  it seems pertinent to ask whether these exact results in field theories can teach us something about the holographic M/string theory duals, beyond the supergravity regime.

There have been a number of works trying to use the localization of Wilson loops in four dimensional ${\cal N}=2$ Yang Mills theories to probe putative string duals \cite{Rey:2010ry, Passerini:2011fe, Fraser:2011qa}. This is a potentially very exciting line of research, as it may reveal properties of holographic pairs that have not been fully established to date. In this work we will take a slightly different route, by applying localization to probe a known example of holographic duality. We will consider ${\cal N}=4$ SYM with gauge group $G=SO(N),Sp(N)$, which is dual to type IIB string theory compactified on $AdS_5\times \bR\bP^5$ with various choices of discrete torsion \cite{Witten:1998xy}\footnote{The precise statement is actually more subtle: given a Lie algebra ${\mathfrak g}$, there is a variety of Lie groups $G$ associated to it, and all of them define different gauge theories. These gauge theories have the same correlators of local operators, but differ in the spectrum of non-local operators \cite{Aharony:2013hda}. In the case of ${\cal N}=4$ SYM, theories with the same ${\mathfrak g}$ and different $G$ each have their own holographic dual, differing by a choice of quantization of certain topological term in the type IIB action \cite{Aharony:1998qu}. We are grateful to Ofer Aharony for clarifying correspondence on this point.}. This duality is closely related to the original proposal for $G=SU(N)$, but it displays a number of novel features, related to the presence of non-orientable surfaces in the $1/N$ expansion of the field theories, or equivalently to the existence of homologically non-trivial non-orientable subvarieties in the gravity background. Our aim is to explore some of these features at finite $g_s$ and $\alpha'/R^2$, taking advantage of the possibility of computing exactly the vev of certain Wilson loop operators for these field theories. While our focus is on non-local operators, the physics of local operators of these field theories at finite N has been explored in \cite{Caputa:2013vla}.

Our first task will be to compute the vev of $1/2$-BPS circular Wilson loops in various representations, for Euclidean ${\cal N}=4$ SYM with gauge groups $G=SO(N),Sp(N)$. Even before we start thinking about holography, the evaluation of these vevs has interesting applications within field theory. For instance, for $G=U(N),SU(N)$, they immediately allow us to compute the Bremsstrahlung functions for the corresponding heavy probes, using the relation \cite{Correa:2012at}
\begin{equation}
B(\lambda,N)_{\cal R}=\frac{1}{2\pi^2}\lambda\partial_\lambda\hbox{ log}\vev{W_{\cal R}}
\end{equation}
valid for any representation ${\cal R}$. These Bremsstrahlung functions in turn completely determine various quantities of physical interest, like the total radiated power \cite{Correa:2012at, Fiol:2012sg} and the momentum fluctuations of the corresponding accelerated probe \cite{Fiol:2013iaa}. These vevs also determine  the exact change in the entanglement entropy of a spherical region when we add a heavy probe \cite{Lewkowycz:2013laa}\footnote{It is worth keeping  in mind that for the computation of the entanglement entropy \cite{Lewkowycz:2013laa}, it is convenient to use a normalization of the Wilson loops different from the one used in this work.}. Finally, they can also be used to carry out detailed tests of S-duality in ${\cal N}=4$ SYM \cite{Gomis:2009ir}. 

The technical computation of these vevs is quite similar to the ones performed for unitary groups, and amounts to introducing a convenient set of orthogonal polynomials to carry out the matrix model integrals. In fact, since for all Lie algebras ${\mathfrak g}$ the matrix model is Gaussian, the relevant orthogonal polynomials are Hermite polynomials, and the computation of vevs ends up amounting to the evaluation of matrix elements for a $N$-fermion state of the one-dimensional harmonic oscillator, 
\begin{equation}
\vev{W}=\frac{\vev{\Psi_{\mathfrak{g}}|W|\Psi_{\mathfrak{g}}}}{\vev{\Psi_{\mathfrak{g}}|\Psi_{\mathfrak{g}}}}
\label{vevslater}
\end{equation}
the only difference being the parity of the one-fermion states involved: for $\mathfrak{su(n)}$, $\left|\Psi\right>$ is built by filling the first $N$ eigenstates of a harmonic oscillator, for $\mathfrak{so(2n)}$ filling the first $N$ even states and for $\mathfrak{g =so(2n+1),sp(n)}$ the first $N$ odd eigenstates \cite{Myers:1990pp, Bilal:1990yy, Caputa:2013vla}. The computations are straightforward, and reveal exact relations among various vevs. For Wilson loops in the respective fundamental representations we find that 
\begin{equation}
\vev{W(g)}_{\substack{SO(2N)\\ Sp(N)}}=\vev{W(g)}_{U(2N)}\mp \frac{1}{2}\int_0^g dg' \;  \vev{W(g')}_{U(2N)}
\label{exactrelint}
\end{equation}
where $g=\lambda/4N$. This in an exact relation, valid for any value of $\lambda$ and $N$.

Once we have obtained these exact field theory results, we shift gears towards string theory. In the past, the exact computation of circular Wilson loops of ${\cal N}=4$ SU(N) SYM has been used for precision tests of AdS/CFT \cite{Drukker:2005kx, Giombi:2006de, Fiol:2011zg}. Our attitude in the present work will be different, we will take for granted the holographic duality, and we aim to use the exact field theory results to learn about string theory on $AdS_5\times \bR\bP^5$. Our first observation actually doesn't even rely on the actual computation of the vevs of Wilson loops, it can be made just by noticing that for $SU(N)$, the $N$-fermion state $\left|\Psi\right>$ in (\ref{vevslater}) is the groundstate of the fermionic system dual to the LLM sector \cite{Lin:2004nb} of $AdS_5\times S^5$. We use this observation to revisit the question \cite{Mukhi:2005cv} of what is the analogue of the LLM sector for type IIB on $AdS_5\times \bR\bP^5$, and argue that it is given by geometries built out of fermions whose wavefunctions have fixed parity, even for $SO(2N)$ and odd for $SO(2N+1),Sp(N)$. In this latter case, those are the wavefunctions of the half harmonic oscillator \cite{Mukhi:2005cv}. Still in the LLM sector, we point out that the absence or presence of discrete torsion in the gravity dual correlates with the sign of the one-fermion Wigner quasi-distribution at the origin of phase space.

Another aspect of the holographic duality where we can put our exact results to work is perturbative string theory around $AdS_5 \times \bR \bP^5$. The idea is not new: consider the vev of the circular Wilson loop in the fundamental representation of $SU(N)$, which is known exactly \cite{Drukker:2000rr}; in principle, string perturbation theory ought to reproduce the $1/N$ expansion of this vev by world-sheet computations at arbitrary genus on $AdS_5\times S^5$. In practice, these world-sheet computations are currently well out of reach. We would like to claim that some of our results for $G=SO(N), Sp(N)$ might have a better chance of being reproduced by direct world-sheet arguments than those of $G=SU(N)$. To see why, let's recall some generic facts about the large N expansion of gauge theories. In this limit, Feynman diagrams rearrange themselves in a topological expansion of two-dimensional surfaces, weighted by $N^\chi$, where $\chi$ is the Euler characteristic of the surface, namely,
$$
\chi=-2h+2-c-b
$$
for a surface with $h$ handles, $c$ crosscaps and $b$ boundaries. For a $U(N),SU(N)$ field theory with all the fields in the adjoint, gauge invariant quantities admit a $1/N^2$ expansion (rather than $1/N$) as befits orientable surfaces. For instance, for the vev of the circular Wilson loop in the fundamental representation of $U(N)$ the relevant world-sheets have a single boundary and an arbitrary number of handles, and in \cite{Drukker:2000rr} it was explicitly shown that this vev admits a $1/N^2$ expansion. On the other hand, it is well-known that the $1/N$ expansion of field theories with $G=SO(N),Sp(N)$ contains both even and odd powers of $1/N$ \cite{Cicuta:1982fu}, signaling the presence of non-orientable surfaces\footnote{ See \cite{Naculich:1993ve} for the $1/N$ expansion of 2d Yang-Mills theory with $G=SO(N),Sp(N)$.}. On general grounds, as discussed in detail below, we can classify the world-sheets as having an arbitrary number of handles, and zero, one or two crosscaps. However, a closer inspection of eq. (\ref{exactrelint}) reveals that in a $1/N$ expansion, the first term of the RHS corresponds to orientable world-sheets, while the second one to world-sheets with a single crosscap. We thus learn that, for these quantities, the contribution from world-sheets with a single crosscap is given by an integral of the contribution from orientable world-sheets, while world-sheets with two cross-caps don't contribute. These two features are peculiar to the very specific vevs we have considered. Nevertheless, since they have been derived from exact field theory relations, before actually carrying out the $1/N$ expansion, it is conceivable that they could be deduced in string theory by symmetry arguments, without having to carry out the world-sheet computations.

The structure of the paper is as follows. In section 2 we define the field theory quantities we want to evaluate, and recall that thanks to localization, they boil down to matrix model computations. We then compute the vev of circular Wilson loops for various gauge groups and representations. In section 3 we discuss implications for string theory of the computations presented in the previous section. Some very basic facts about classical simple Lie algebras that we use in the main text are collected in appendix A, while in appendix B we present an alternative derivation of some of the results obtained in section 3.

\section{Computations}
This section is entirely devoted to the computation of vevs of circular Wilson loops in ${\cal N}=4$ SYM, leaving for the next section the discussion of the implications of the results found here. Technically, the evaluation of these vevs of Wilson loops is possible since they localize to a computation in a Gaussian matrix model \cite{Erickson:2000af,Drukker:2000rr, Pestun:2007rz}, with matrices in the Lie algebra ${\mathfrak g}$. To carry out the remaining integrals, we resort to the well-known technique of orthogonal polynomials (see \cite{Di Francesco:1993nw} for reviews). Besides the specific results we find, the main point to keep in mind from this section is that for all classical Lie algebras, the orthogonal polynomials are Hermite polynomials, the main difference being the restrictions on their parity. Namely, for the A, B/C and D series, the Hermite polynomials that play a role have unrestricted, odd and even parity, respectively. This observation will become important in the next section. 

The field theory quantities we want to compute are vevs of locally BPS Wilson operators. These Wilson loops are determined by a representation ${\cal R}$ of the gauge group $G$ and a contour ${\cal C}$,
\begin{equation}
W_{\cal R}[{\cal C}]=\frac{1}{\hbox{dim }{\cal R}}\hbox{Tr}_{\mathcal R}{\cal P}\hbox{exp }
\left(i \int_{\cal C} (A_\mu \dot x^\mu +|\dot x|\Phi_i \theta^i)ds \right)
\end{equation}
We have fixed the overall normalization of the Wilson loop by the requirement that at weak coupling, $\vev{W_{\cal R}}=1+{\cal O}(g)$. We will be interested in the case when the signature is Euclidean and the contour is a circle. These Wilson loops are 1/2 BPS and remarkably the problem of the evaluation of their vev localizes to a Gaussian matrix model computation  \cite{Erickson:2000af,Drukker:2000rr, Pestun:2007rz},
$$
\vev{W}_{\cal R}=\frac{1}{\hbox{dim }{\cal R}}
\frac{\int_{\mathfrak g} dM e^{-\frac{1}{2g} \hbox{tr }M^2} \; \hbox {Tr}_{\cal R} e^M}{\int_{\mathfrak g} dM e^{-\frac{1}{2g} \hbox{tr }M^2}}
$$
where the integrals are over the Lie algebra $\mathfrak g$ and $g=\lambda/4N$. These integrals can be reduced to integrals over the Cartan subalgebra $\mathfrak h$ (see \cite{Gomis:2009ir} for details), and one arrives at
\begin{equation}
\vev{W}_{\cal R}=\frac{1}{\hbox{dim }{\cal R}}
\frac{\int_{\mathfrak h} dX  \Delta(X)^2 e^{-\frac{1}{2g} \hbox{tr }X^2} \; \hbox {Tr}_{\cal R} e^X}{\int_{\mathfrak h} dX  \Delta(X)^2 e^{-\frac{1}{2g} \hbox{tr }X^2}}
\label{intecartan}
\end{equation}
where the Jacobian $\Delta(X)^2$ is given by a product over positive roots of the algebra,
\begin{equation}
\Delta(X)^2=\prod_{\alpha >0} \alpha(X)^2
\label{bigdelta}
\end{equation}
As in  \cite{Gomis:2009ir}, it is convenient to write the insertion of the Wilson loop as a sum over the weights of the representation,
\begin{equation}
\hbox{Tr}_{\cal R} e^X=\sum_{v\in \Omega({\cal R})} n(v) e^{v(x)}
\label{chartrace}
\end{equation}
where $\Omega ({\cal R})$ is the set of weights $v$ of the representation ${\cal R}$, and $n(v)$ the multiplicity of the weight. Now that we have introduced the matrix integrals that we want to compute let's very briefly recall the technique we will use to solve them, the method of orthogonal polynomials. Given a potential $W(x)$, we can define a family of orthogonal polynomials $p_n(x)$ satisfying
$$
\int_\infty^\infty dx \; p_m(x)p_n(x) e^{-\frac{1}{g}W(x)} =h_n \delta_{mn}
$$
We will choose these polynomials to be monic, namely $p_n(x)=x^n+{\cal O}(x^{n-1})$. More precisely, in all the cases in this work, the potential is $W(x)=\frac{1}{2}x^2$, and the orthogonal polynomials are Hermite polynomials,
\begin{equation}
p_n(x)=\left(\frac{g}{2}\right)^{\frac{n}{2}}H_n\left(\frac{x}{\sqrt{2g}}\right)
\label{thepoly}
\end{equation}
so in our conventions
$$
h_n=g^n \sqrt{2\pi g} \; n!
$$
For future reference, recall that these polynomials have well-defined parity, $p_n(-x)=(-1)^n p_n(x)$. The key point is that in all cases we will encounter in this work, the Jacobian $\Delta(X)^2$ in (\ref{bigdelta}) can be substituted by the square of a determinant of orthogonal polynomials. Once we perform this substitution, we expand the determinants using Leibniz formula and carry out the resulting integrals. Note also that the determinant of orthogonal polynomials combined with the Gaussian exponent is (up to a normalization factor) the Slater determinant that gives the wave-function of an $N$-fermion state,
 $$
 \left|\Psi_N(x_1,\dots,x_N)\right >=C|H_i(x_j)e^{-\frac{1}{4g}x_j^2}|
$$
so in all cases the computations we perform can be thought of as normalized matrix elements for certain $N$-fermion states
\begin{equation}
\vev{{\cal O}}_{mm}=\frac{\vev{\Psi_N|{\cal O}|\Psi_N}}{\vev{\Psi_N|\Psi_N}}
\label{slatervev}
\end{equation}
where the specific $\left|\Psi_N\right >$ depends on the algebra $\mathfrak{g}$. For $G=SO(N),Sp(N)$, these Slater determinants involving one-fermion wavefunctions of definite parity also appear in the study of certain local operators \cite{Caputa:2013vla}.

Having reviewed all the ingredients we now turn to some explicit computations. We use some very basic facts of classical Lie algebras, that we have collected in appendix A.

\subsection{$\mathfrak {su(n)}$ }
This case is the best studied one, corresponding to the familiar Hermitian matrix model. It is customary to work with $U(N)$, and we will do so in what follows; the modification needed when dealing with $SU(N)$ is mentioned below. While none of the results recalled here are new, having them handy will be helpful in what follows. In this case, the Jacobian  (\ref{bigdelta}) is
$$
\prod_{\alpha >0} \alpha(X)^2=\prod_{1\leq i<j\leq N} |x_i-x_j|^2
$$
This Vandermonde determinant can be traded by a determinant of polynomials, which due to the Gaussian potential is convenient to choose to be the first N Hermite polynomials (\ref{thepoly}),
\begin{equation}
\prod_{1\leq i<j\leq N} |x_i-x_j|=|p_{i-1}(x_j)|
\label{vandersu}
\end{equation}
The partition function can be computed using (\ref{vandersu})
\begin{align}
\nonumber
{\cal Z}&=\int _{-\infty}^\infty dx_1 \dots \int _{-\infty}^\infty dx_N \prod _{1\leq i<j\leq N} |x_i-x_j|^2 \; e^{-\frac{1}{2g} (x_1^2+\dots +x_N^2)} = \\ 
  & =\int _{-\infty}^\infty dx_1 \dots \int _{-\infty}^\infty dx_N \; |p_{i-1}(x_j)|^2 \; e^{-\frac{1}{2g} (x_1^2+\dots +x_N^2)}  =N!\prod_{i=0}^{N-1} h_i
 \label{partfunsu}
\end{align}
In the last step we used the following integral of Hermite polynomials \cite{gradshteyn}, that we will apply repeatedly in this work,
\begin{equation}
\int_{-\infty}^{\infty} H_m(x)H_n(x) e^{-(x-y)^2} dx =  2^n \sqrt{\pi} m! \, y^{n-m} L_m^{n-m}(-2y^2) \hspace{.8cm} n\geq m
\label{theinte}
\end{equation}
where $L_n^\alpha(x)$ are generalized Laguerre polynomials.

Let's recall briefly the computation of Wilson loops. Consider first the Wilson loop in the fundamental representation\footnote{A Lie algebra of rank $r$ has $r$ fundamental weights, which are the highest weights of the $r$ fundamental representations. In Physics 'fundamental representation' often refers to the representation with highest weight $w_1$}. The new integral to compute is 
\begin{align*}
& \int _{-\infty}^\infty dx_1 \dots \int _{-\infty}^\infty dx_N \prod _{1\leq i<j\leq N} |x_i-x_j|^2 \;  \left(e^{x_1}+\dots+ e^{x_N}\right) e^{-\frac{1}{2g} (x_1^2+\dots +x_N^2)} =\\
=N &  \int _{-\infty}^\infty dx_1 \dots \int _{-\infty}^\infty dx_N  |p_{i-1}(x_j)|^2 \; e^{x_1} e^{-\frac{1}{2g} (x_1^2+\dots +x_N^2)} 
\end{align*}
where we already used (\ref{vandersu}). Now applying (\ref{theinte}) and recalling (\ref{partfunsu}) we arrive at \cite{Drukker:2000rr}
\begin{equation}
\vev{W(g)}_{U(N)}=\frac{1}{N}\sum_{k=0}^{N-1}L_{N-1}(-g)e^{\frac{g}{2}}=\frac{1}{N}L_{N-1}^1(-g) e^{\frac{g}{2}}
\label{exactfund}
\end{equation}
The remaining U(N) fundamental representations are the k-antisymmetric representation. The exact vevs of the corresponding Wilson loops were computed in \cite{Fiol:2013hna}. In order to evaluate vevs of Wilson loops for $SU(N)$, we have to modify the insertion to \cite{Drukker:2000rr, Okuda:2008px}
$$
\hbox{Tr}_{\cal R} e^X \rightarrow   e^{-\frac{|{\cal R}|}{N} Tr X} \; \hbox{Tr}_{\cal R} e^X
$$

\subsection{$\mathfrak {so(2n)}$}
The Jacobian $\Delta(X)^2$ for these algebras is
$$
\prod_{\alpha >0} \alpha(X)^2= \prod_{1\leq i<j\leq N} |x_i^2-x_j^2|^2
$$
The key argument to evaluate all the integrals we will encounter in this case rests on two facts: first, the expression above for $\Delta^2(X)$ is a Vandermonde determinant of $\{x_i^2\}$ and second, even polynomials $p_{2i}(x)$ involve only even powers of $x$, so it is possible to replace
\begin{equation}
\prod _{1\leq i<j\leq N} |x_i^2-x_j^2|^2=|p_{2(i-1)}(x_j)|^2
\label{jacobso}
\end{equation}
It is worth pointing out that while for $\mathfrak {g=su(n)}$, the Hermite polynomials that appear in eq. (\ref{vandersu}) correspond to the first $N$ eigenstates of the harmonic oscillator, for $\mathfrak{so(2n)}$ what appears in (\ref{jacobso}) are the first $N$ even eigenstates, so only those will contribute to the computation of the partition function and the vev of Wilson loops. Let's start by evaluating the partition function of the corresponding matrix model,
$$
{\cal Z}  =  \int _{-\infty}^\infty dx_1 \dots \int _{-\infty}^\infty dx_N \prod _{1\leq i<j\leq N} |x_i^2-x_j^2|^2 e^{-\frac{1}{2g} (x_1^2+\dots +x_N^2)} 
$$
Performing the substitution (\ref{jacobso}), we arrive at
\begin{equation}
{\cal Z}=N! \prod_{i=0}^{N-1}h_{2i}
\label{partfunso}
\end{equation}
Let's now compute the vev of Wilson loops in various fundamental representations. As a first example, let's choose the representation with highest weight $w_1$. The $2N$ weights of this representation are $e_i$ and $-e_i$ for $i=1,\dots,N$. After diagonalization, the matrix model that computes the vev of the Wilson loop is
$$
\vev{W(g)}_{SO(2N)}=\frac{1}{{\cal Z}}\int _{-\infty}^\infty dx_1 \dots \int _{-\infty}^\infty dx_N \prod _{1\leq i<j\leq N} |x_i^2-x_j^2|^2 \frac{e^{x_1}+e^{-x_1}}{2} e^{-\frac{1}{2g} (x_1^2+\dots +x_N^2)}
$$
Performing the substitution (\ref{jacobso}), taking into account (\ref{partfunso}) and using (\ref{theinte}) we arrive at
\begin{equation}
\vev{W(g)}_{SO(2N)}=\frac{1}{N}\sum_{k=0}^{N-1}L_{2k}(-g) e^{g/2}
\label{exactso}
\end{equation}
Let's now compute the vev of a Wilson loop in a spinor representation\footnote{In $AdS_5\times \bR \bP^5$, these Wilson loops are dual to a D5-brane wrapping $\bR \bP^4 \subset \bR \bP^5$ \cite{Witten:1998xy}.}. The spinor representation with highest weight $w_{N-1}$ has weights of the form 
$$
\frac{1}{2}\left(\pm e_1\pm e_2 \pm \dots \pm e_N\right)
$$
with an odd number of minus signs, while the representation with highest weight $w_N$ has weights with an even number of minus signs. Let's focus on the representation with highest weight $w_N$,
$$
\vev{W}_{w_N}=\frac{1}{{\cal Z}} \frac{1}{2^{N-1}} \int _{-\infty}^\infty dx_1 \dots \int _{-\infty}^\infty dx_N \prod _{1\leq i<j\leq N} |x_i^2-x_j^2|^2 \sum_{\substack{\{s_i=\pm\} \\ \prod_i s_i=1}}
e^{\frac{1}{2}\left( s_1x_1+\dots +s_Nx_N\right)}e^{-\frac{1}{2g} (x_1^2+\dots +x_N^2)}
$$
For each $s_i=-$, we change variables $\tilde x_i=-x_i$, and deduce that all $2^{N-1}$ terms contribute the same to the full integral,
\begin{align*}
\vev{W}_{w_N} & =\frac{1}{{\cal Z}} \int _{-\infty}^\infty dx_1 \dots \int _{-\infty}^\infty dx_N \prod _{1\leq i<j\leq N} |x_i^2-x_j^2|^2 e^{\frac{1}{2}\left(x_1+\dots +x_N\right)}e^{-\frac{1}{2g} (x_1^2+\dots +x_N^2)} = \\
 &=\frac{1}{{\cal Z}} \int _0^\infty dx_1 \dots \int _0^\infty dx_N \prod _{1\leq i<j\leq N} |x_i^2-x_j^2|^2 \prod_{i=1}^N \left(e^{\frac{x_i}{2}}+e^{-{\frac{x_i}{2}}}\right) e^{-\frac{1}{2g} (x_1^2+\dots +x_N^2)}
\end{align*}
Now the remaining integrals can be solved as before. After using the substitution (\ref{jacobso}) the details are quite similar to the computation of the vev of Wilson loops in antisymmetric representations of $U(N)$ \cite{Fiol:2013hna}, so we will skip the details and just present the final result. Define the $N\times N$ matrix $D_{ij}$, with entries involving generalized Laguerre polynomials $L_n^\alpha(x)$,
$$
D_{ij}=L_{2i-2}^{2j-2i}(-g/4)e^{g/8}
$$
Then, the vev of the Wilson loop in the $w_N$ representation is
$$
\vev{W}_{w_N}=|D_{ij}|
$$
Expanding the determinant, and following identical steps as those presented in \cite{Fiol:2013hna}, we can rewrite this vev as
$$
\vev{W}_{w_N}=P_N(g)e^{\frac{\lambda}{32}}
$$
where $P_N(g)$ is a polynomial in $g$ of degree $N(N-1)/2$ that can be written as a sum involving ordered N-tuples,
$$
P_N(g)=\sum_{0\leq \tau_1<\tau_2<\dots \tau_N\leq 2N-2} \prod_{m=1}^N \frac{\tau_m!}{(2m-2)!} \left|{2i \choose \tau_j}\right|^2 \left(\frac{g}{4}\right)^{N(N-1)-\sum_{m=1}^N \tau_m}
$$
The other spinor representation, with highest weight $w_{N-1}$, has weights with an odd numbers of minus signs, but applying the same change of variables $\tilde x_i=-x_i$ to all minus signs, we immediately arrive at the same integral as before, so we conclude that both vevs are the same,
$$
\vev{W}_{w_{N-1}}=\vev{W}_{w_N}
$$

\subsection{$\mathfrak {sp(n)}$}
In this case we have
$$
\prod_{\alpha >0} \alpha(X)^2= \prod_{1\leq i<j\leq N} |x_i^2-x_j^2|^2 \prod_{i=1}^N x_i^2
$$
Again, since odd Hermite polynomials involve only odd powers of $x$, it is possible to substitute the Jacobian by the square of a determinant of orthogonal polynomials
\begin{equation}
\prod _{1\leq i<j\leq N} |x_i^2-x_j^2|^2 \prod_i x_i^2 =|p_{2i-1}(x_j)|^2
\label{jacobsp}
\end{equation}
where now the polynomials that appear correspond to the first $N$ odd eigenstates of the harmonic oscillator. The partition function can be readily computed
\begin{equation}
{\cal Z}=N! \prod_{i=1}^N h_{2i-1}
\label{partisp}
\end{equation}
Let's now turn to the computation of Wilson loops. Let's compute for example the vev of the Wilson loop in the representation with highest weight $w_1$. The weights are $e_i$ and $-e_i$ for $i=1,\dots,N$. After diagonalization, the matrix model that computes the vev of the Wilson loop is
$$
\vev{W(g)}_{Sp(N)}=\frac{1}{{\cal Z}}\int _\infty^\infty dx_1 \dots  dx_N \prod _{1\leq i<j\leq N} |x_i^2-x_j^2|^2 \prod_i x_i^2 \frac{e^{x_1}+e^{-x_1}}{2} e^{-\frac{1}{2g} (x_1^2+\dots x_N^2)}
$$
Using the substitution (\ref{jacobsp}), taking into account (\ref{partisp}) and (\ref{theinte}), we arrive at
\begin{equation}
\vev{W(g)}_{Sp(N)}=\frac{1}{N}\sum_{k=0}^{N-1}L_{2k+1}(-g) e^{g/2}
\label{exactsp}
\end{equation}

\subsection{$\mathfrak{so(2n+1)}$}
The Jacobian is the same as for $\mathfrak{sp(n)}$, so it admits the same replacement
$$
\prod_{\alpha >0} \alpha(X)^2= \prod_{1\leq i<j\leq N} |x_i^2-x_j^2|^2 \prod_{i=1}^N x_i^2
$$
The partition function is essentially the same as for $\mathfrak{sp(n)}$, eq. (\ref{partisp}). Let's compute some vevs of Wilson loops.  As a first example, consider the representation with highest weight $w_1$. The weights of this representation are $e_i$ and $-e_i$ for $i=1,\dots,N$ plus the zero weight. After diagonalization, the matrix model that computes the vev of the Wilson loop is
\begin{align*}
\vev{W(g)}_{SO(2N+1)}=& \frac{1}{2N+1}\frac{1}{{\cal Z}}\int _\infty^\infty dx_1 \dots  dx_N \prod _{1\leq i<j\leq N} |x_i^2-x_j^2|^2 \prod_i x_i^2 \\
& \left(1+e^{x_1}+e^{-x_1}+\dots +e^{x_N}+e^{-x_N}\right) e^{-\frac{1}{2g} (x_1^2+\dots +x_N^2)}
\end{align*}
Now, the measure is the same as for $\mathfrak{sp(n)}$, so the same substitution (\ref{jacobsp}) works here, and we arrive at
$$
\vev{W(g)}_{SO(2N+1)}=\frac{1}{2N+1}\left(1+2\sum_{k=0}^{N-1}L_{2k+1}(-g) e^{g/2}\right)
$$
For the spinor representation of $\mathfrak{so(2n+1)}$, the computation proceeds along the same lines as for the spinor representations of $\mathfrak{so(2n)}$. Let's just quote the result; define the $N\times N$ matrix 
$$
B_{ij}=L_{2i-1}^{2j-2i}(-g/4)e^{g/8}
$$
Then
$$
\vev{W}_{w_N}=|B|
$$

\section{Implications}
In the last section we have computed the exact vev of circular Wilson loops of ${\cal N}=4$ SYM, for various representations of different gauge groups. In what follows, we are going to discuss some features and implications of the results we have obtained. Our main interest is trying to derive lessons for the holographic duals of these gauge theories. 

The string dual of ${\cal N}=4$ SYM with gauge group $SU(N)$ is of course type IIB string theory on $AdS_5\times S^5$. For ${\cal N}=4$ with gauge groups $SO(N), Sp(N)$ one can argue for the string duals as follows \cite{Witten:1998xy}. Start by placing N parallel D3-branes at an orientifold three-plane. Taking the near horizon limit, the theory on the world-volume of the D3-branes becomes ${\cal N}=4$ SYM with gauge group $SO(N),Sp(N)$ while the supergravity solution becomes $AdS_5\times \bR \bP^5$ (Recall that $\bR \bP^5$ is $S^5/\bZ_2$ with $\bZ_2$ acting as $x_i\sim -x_i$). This orientifold is common to all the holographic duals for $SO(2N),SO(2N+1),Sp(N)$. The additional ingredients  that discriminate among these duals are the possible choices of discrete torsion. Let's recall very briefly the identification of these supergravity duals, referring the interested reader to \cite{Witten:1998xy} for the detailed derivation.  In the presence of the orientifold, the B-fields $B_{NS}$ and $B_{RR}$ become twisted two-forms. The possible choices of discrete torsion for each of them are classified by  $H^3(\bR \bP^5,\tilde \bZ)=\bZ_2$, so calling $\theta_{NS}$ and $\theta_{RR}$ these two choices, there are all in all four possibilities. Using the transformation properties of ${\cal N}=4$ SYM with different gauge groups under Montonen-Olive duality, it is possible to identify the choices of discrete torsion for the respective gravity duals. The choices $(\theta_{NS},\theta_{RR})=(0,0), (0,1/2),(1/2,0), (1/2,1/2)$ correspond to the gauge groups $SO(2N),SO(2N+1),Sp(N), Sp(N)$ respectively\footnote{These last two $Sp(N)$ theories differ by their value of the $\theta$ angle.}. 

\subsection{The LLM sector}
The first aspect of the holographic duality that we are going to consider is the analogue of the LLM geometries \cite{Lin:2004nb} in $AdS_5\times \bR\bP^5$. Let's recall briefly that LLM \cite{Lin:2004nb} constructed an infinite family of ten dimensional IIB supergravity solutions, corresponding to the backreaction of 1/2 BPS states associated to chiral primary operators built out of a single chiral scalar field. These ten dimensional solutions are completely determined by a single function $u(x_1,x_2)$ of two spacetime coordinates. For regular solutions, this function can take only the values $u(x_1,x_2)=0,1$ defining a "black-and-white" pattern on the $x_1,x_2$ plane\footnote{This function $u(x_1,x_2)$ is related to the function $z(x_1,x_2)$ of the original paper \cite{Lin:2004nb} by $u=1/2-z$.}. On the field theory side, the dynamics of this sector of operators of ${\cal N}=4$ SU(N) SYM is controlled by the matrix quantum mechanics of N fermions on a harmonic potential \cite{Corley:2001zk, Berenstein:2004kk}. The one-fermion phase space $(q,p)$ gets identified with the $(x_1,x_2)$ plane displaying the "black-and-white" pattern. In particular, the ground state of the system is given by filling the first N states of the harmonic oscillator; in the one-fermion phase space, this corresponds to a circular droplet, which in turn is the pattern giving rise to the $AdS_5\times S^5$ solution in supergravity. The fermion picture can be inferred directly from the supergravity solutions \cite{Mandal:2005wv,Grant:2005qc}.

This is the LLM sector of the duality between type IIB on $AdS_5\times S^5$ and ${\cal N}=4$ SU(N) SYM. What is the similar sector for ${\cal N}=4$ SYM with $G=SO(N),Sp(N)$ ? We are going to propose an answer motivated by the fact that the groundstate of the LLM sector for SU(N) is precisely the N-fermion state $\left |\Psi_N\right >$ that appears in the matrix model that computes Wilson loops, eq. (\ref{slatervev}). We then propose that for the other classical Lie algebras, it also holds that the corresponding $\left |\Psi_{\mathfrak {g}}\right>$ in eq. (\ref{slatervev}) is the groundstate of the fermionic system dual to the LLM sector. We can imagine starting with the matrix model for $U(2N)$, so in the ground state the fermions fill up the first 2N energy levels, and then the orientifold projects out either the even or odd parity eigenstates, depending on the gauge group we consider. The LLM sectors are certainly richer than just the groundstate: they are given by a matrix quantum mechanics that allows for excitations. Our complete proposal is that the full LLM sectors are given by {\it any} N fermion state built from one-fermion eigenstates of fixed parity: even parity for $SO(2N)$ and odd parity for $SO(2N+1),Sp(N)$,
\begin{equation}
\psi(-x)=(-1)^s \psi(x)
\label{waveproj}
\end{equation}
where $s=0,1$ depending on the gauge group. This picture is especially easy to visualize for $SO(2N+1),Sp(N)$ since in these cases we are keeping odd-parity eigenstates, which are the eigenstates of an elementary problem in 1d quantum mechanics: the "half harmonic oscillator" where we place an infinite wall at the origin of a harmonic oscillator potential. This identification between the orientifold in $AdS_5\times \bR \bP^5$ and the projection from the harmonic oscillator to the half harmonic oscillator was pointed out in \cite{Mukhi:2005cv}, where it was suggested to hold for any $SO(N),Sp(N)$ group. According to our argument, this identification holds for $SO(2N+1),Sp(N)$, but it does not for $SO(2N)$, since in this case the states preserved by the orientifold action are the even parity ones. 

We can formalize this identification as follows. In \cite{Mukhi:2005cv} it was argued that the orientifold projection acts in the $(x_1,x_2)$ plane of LLM geometries as $(x_1,x_2)\sim (-x_1,-x_2)$. Since the $(x_1,x_2)$ plane is identified with the one-fermion phase space, this identification amounts to implementing a parity projection in phase space. To do so, one can define \cite{Royer:1977zz} the following parity operator in phase space
\begin{equation}
\Pi_{q,p}=\int _{-\infty}^\infty ds \; e^{-2ips/\hbar} \; \left|q-s\right >\left <q+s\right|
\label{parityop}
\end{equation}
and the projectors
$$
P_{q,p}^\pm=\frac{1}{2}\left(1\pm \Pi_{q,p}\right)
$$
In particular, $\Pi_{(0,0)}$ is the parity operator about the origin of phase space: it changes $\psi(q)$ into $\psi(-q)$ and $\hat \psi(p)$ into $\hat \psi(-p)$, so the similarity with the orientifold action is apparent. The projectors $P_{0,0}^\pm$ project on the space of wavefunctions symmetric or antisymmetric about the origin, and the orientifold projection amounts to keeping one of these subspaces.

Going forward with the argument, we note that $s=0,1$ in eq. (\ref{waveproj}), depending on the absence or presence of discrete torsion. We want to provide a new perspective on this discrete torsion, from the phase space point of view. We start by recalling that the function $u(x_1,x_2)$ is identified with the phase space density $u(p,q)$ of one of the fermions in the system of N fermions in a harmonic potential. To go beyond a purely classical description, one can consider a number of phase space quasi-distributions that replace the phase space density, as has been discussed in the LLM context in \cite{Takayama:2005yq, Balasubramanian:2005mg}. One particular such distribution is the Wigner distribution, defined as the Wigner transform of the density matrix,
$$
{\cal W}(p,q)=\frac{1}{\pi \hbar}\int_{-\infty}^{\infty}dy \; e^{2ipy/\hbar} \left<q-y|\hat \rho|q+y\right >
$$
A salient feature of Wigner quasi-distributions is that they are not positive definite functions over phase space. For instance, if we consider a given eigenstate $\left|n\right>$ of the harmonic oscillator, the corresponding Wigner distribution is given again by a Laguerre function \cite{Takayama:2005yq, Balasubramanian:2005mg}\footnote{At this time, we regard the fact that Laguerre functions appear both in the vevs of circular Wilson loops and in Wigner distributions as merely fortuitous. In particular, note that the vevs of Wilson loops have negative argument, while for Wigner distributions the argument is positive.}
$$
{\cal W}_n(p,q)= \frac{(-1)^n}{\pi \hbar} L_n\left(2\frac{q^2+p^2}{\hbar}\right)e^{-\frac{q^2+p^2}{\hbar}}
$$ 
In particular, for the eigenstate $\left|n\right >$, at the origin of phase space we have 
$$
{\cal W}_n(0,0)=(-1)^n\frac{1}{\pi \hbar}
$$
so it can have either sign. More generally, the Wigner quasi-distribution is the expectation value of the parity operator defined in (\ref{parityop}) \cite{Royer:1977zz}
$$
{\cal W}(p,q)=\frac{1}{\pi \hbar} \vev{\Pi_{p,q}}
$$
and in particular
$$
{\cal W}(0,0)=\frac{1}{\pi \hbar}\vev{\Pi_{0,0}}
$$
so it is clear that the sign of ${\cal W}(0,0)$ captures the parity properties of the wavefunction with respect to the origin of phase space\footnote{Incidentally, negative values of the Wigner function at the origin of phase space have apparently been measured experimentally for single photon fields   \cite{nogues}.}. For a generic N fermion state with eigenstates $\{j_1,\dots,j_N\}$, the Wigner function is \cite{Takayama:2005yq, Balasubramanian:2005mg},
$$
{\cal W}(p,q)=\frac{1}{\pi \hbar} e^{-(q^2+p^2)/\hbar} \sum_{\{j_i\}} (-1)^{j_i} L_{j_i}\left(\frac{2}{\hbar}(q^2+p^2)\right)
$$
For $G=SO(N),Sp(N)$, the sign $(-1)^{j_i}$ is the same for all states, to it comes out of the sum. In particular, for any N fermion state, at the origin of phase space we get
$$
(-1)^s=\hbox{sign }{\cal W}(0,0)
$$

\subsection{Features of the non-orientable terms}
In the previous section we have computed the vevs of circular Wilson loops for various gauge groups and representations. We now want to present some exact relations among these vevs, as well as their large $N$ expansion, which in principle ought to be reproduced by string theory computations on $AdS_5\times \bR\bP^5$. Before we take a detailed look at the results we have obtained, let's recall briefly some general expectations. In the large $N$ expansion, Feynman diagrams rearrange themselves in a topological expansion in terms of two-dimensional surfaces. Each surface is weighted by $N^\chi$, with $\chi$ the Euler characteristic of the surface; for a surface with $h$ handles, $b$ boundaries and $c$ crosscaps, the Euler characteristic is
\begin{equation}
\chi=-2h+2-c-b
\label{eulerchar}
\end{equation}
As a consequence of the classification theorem for closed surfaces, a general non-orientable surface can be thought of as an orientable surface with a number of crosscaps. Furthermore, according to Dycks' theorem, three crosscaps can be traded for a handle and a single crosscap, so we expect three kinds of contributions, coming from world-sheets with an arbitrary number of handles and with zero (i.e. orientable), one or two crosscaps. 

For a $U(N),SU(N)$ theory with all fields in the adjoint representation, the large N expansion of any observable is actually a $1/N^2$ expansion (without odd powers of $1/N$) as it befits an expansion in orientable surfaces. For the vev of a circular Wilson loop of $U(N)$ in the fundamental representation, this $1/N^2$ expansion of the exact result was already carried out in \cite{Drukker:2000rr}\footnote{The surfaces that appear in the $1/N$ expansion of $\vev{W}_{SU(N)}$ have a single boundary and an arbitrary number of handles, so they all have odd Euler characteristic, eq. (\ref{eulerchar}). However, in the normalization for $\vev{W}_{SU(N)}$ followed in \cite{Drukker:2000rr} and in the present work, there is an additional overall $1/N$, so the expansion ends up being in even powers of $N$. At any rate, what is relevant is that the expansion parameter is $1/N^2$ and not $1/N$.}. On the other hand, when $G=SO(N),Sp(N)$, the adjoint representation can be thought of as the product of two fundamental representations (rather than a fundamental times an antifundamental representation as in $U(N)$), so propagators can still be represented by a double line notation, but now without any arrows in the lines \cite{Cicuta:1982fu}. As a result,  the large N expansion of observables for $SO(N),Sp(N)$ theories - even when all fields transform in the adjoint representation - involves both even and odd powers of $1/N$, signaling the appearance of non-orientable surfaces \cite{Cicuta:1982fu}. Furthermore, gauge invariant quantities for $Sp(N)$ are related to those of $SO(2N)$ by the replacement  $N\rightarrow -N$ \cite{mkrtchyan}. Finally, we know that $SO(2N)$ and $Sp(N)$ theories can be obtained from orientifolding $U(2N)$. All in all, these general arguments imply that vevs in the respective fundamental representations of various groups ought to be related by\footnote{Recall that we are normalizing all vevs such that $\vev{W}=1+{\cal O}(g)$. In other normalizations of the vevs of Wilson loops, this equation might involve a different numerical coefficient in front of $\vev{W}_{U(2N)}$.}
\begin{equation}
\vev{W}_{\substack{SO(2N)\\ Sp(N)}}=\vev{W}_{U(2N)}\pm \hbox{unoriented }_{c=1}+\hbox{unoriented }_{c=2}
\label{sospexp}
\end{equation}
where $unoriented$ refers to terms that in the large N limit arrange themselves into non-orientable surfaces with either one or two crosscaps. In the formula above, we have already imposed the relation $Sp(N)=SO(-2N)$, which implies that world-sheets with a single cross-cap  contribute the same for $SO$ and $Sp$ up to a sign, while world-sheets with two cross-caps give the same contribution for the two groups.

We are now going to show that indeed our exact results (\ref{exactso}) and (\ref{exactsp}) follow the pattern expressed in (\ref{sospexp}). In the process, we will furthermore find a couple of features that do not follow from these general arguments.

To obtain the $1/N$ expansion of $\vev{W}_{SO(2N)}$ and $\vev{W}_{Sp(N)}$, we can analyze them separately, following the steps of \cite{Drukker:2000rr}, as we do in the appendix. However, it is much more efficient to consider their sum and their difference, and expand those. Let's start considering the sum. Recalling eq. (\ref{exactfund}), it is immediate that the results we have found, eqs. (\ref{exactso}) and (\ref{exactsp}) satisfy
\begin{equation}
\vev{W(g)}_{SO(2N)}+\vev{W(g)}_{Sp(N)}=2\vev{W(g)}_{U(2N)}
\label{exactsum}
\end{equation}
As for the difference $\vev{W(g)}_{Sp(N)}-\vev{W(g)}_{SO(2N)}$, using properties of the Laguerre polynomials, it is not difficult to prove from the explicit results eqs. (\ref{exactso}) and (\ref{exactsp}) that the following exact relation holds
\begin{equation}
\frac{\partial}{\partial \lambda}\left(\vev{W(g)}_{Sp(N)}-\vev{W(g)}_{SO(2N)}\right)=\frac{1}{4N}\vev{W(g)}_{U(2N)}
\label{exactdiff}
\end{equation}
These last two relations, eqs. (\ref{exactsum}) and (\ref{exactdiff}), can we rewritten in the following suggestive form
\begin{equation}
\vev{W(g)}_{\substack{SO(2N)\\ Sp(N)}}=\vev{W(g)}_{U(2N)}\mp \frac{1}{2}\int_0^g dg' \; \vev{W(g')}_{U(2N)}
\label{exactrela}
\end{equation}
Recall that $\vev{W(g)}_{U(2N)}$ has a expansion in $1/N^{2}$. Furthermore, since $g=\lambda/4N$, the integral brings an extra power of $1/N$. Therefore, equation (\ref{exactrela}) neatly splits the $1/N$ expansions of $\vev{W(g)}_{SO(2N)}$ and $\vev{W(g)}_{Sp(N)}$ into even and odd powers of $1/N$. The $1/N^{2k}$ terms coincide for both vevs and are given $\vev{W(g)}_{U(2N)}$; they correspond to orientable surfaces. Note in particular that since all even powers of $1/N$ come from orientable surfaces, there are no contributions from world-sheets with two crosscaps, as it can be already deduced from eqs. (\ref{sospexp}) and (\ref{exactsum}).

Turning now to the $1/N^{2k+1}$ terms in the expansion of $\vev{W(g)}_{SO(2N)}$ and $\vev{W(g)}_{Sp(N)}$, they come from the integral in eq. (\ref{exactrela}), so it is manifest that they differ just by a sign; this, together with the equality of the even terms in the expansions, proves that indeed $\vev{W(g)}_{Sp(N)}$ can be obtained from $\vev{W(g)}_{SO(2N)}$ by the substitution $N\rightarrow -N$, as it had to happen according to general arguments \cite{mkrtchyan}.

To recapitulate, the $1/N$ expansion of  $\vev{W(g)}_{SO(2N)}$ and $\vev{W(g)}_{Sp(N)}$ could in principle involve contributions from three kinds of surfaces, with zero, one or two crosscaps.
By a mix of generic arguments and exact field theory computations, we have found that for these quantities, and for any number of handles,  contributions from surfaces with one crosscap are given by an integral of the contribution from surfaces without crosscaps, while there is no contribution from surfaces with two crosscaps, eq. (\ref{exactrela}). 

The two features that we have just uncovered for the $1/N$ expansion of  $\vev{W(g)}_{SO(2N)}$ and $\vev{W(g)}_{Sp(N)}$ bear certain resemblance with properties encountered in other instances of $1/N$ expansion of $SO/Sp$ gauge theories. A first example is the computation of the effective glueball superpotential of ${\cal N}=1$ SYM theories with a scalar field in the adjoint, with an arbitrary tree-level polynomial superpotential, ${\cal W}(\Phi)$. Dijkgraaf and Vafa \cite{Dijkgraaf:2002dh} pointed out that for $G=U(N)$, this computation reduces to an evaluation of the planar free energy of a one-matrix model with the matrix model potential given by the tree-level superpotential of the gauge theory. For ${\cal N}=1$ SYM with gauge groups $SO(N),Sp(N)$ the corresponding matrix models are, like in the present work, valued on the Lie algebras  \cite{Ashok:2002bi}. It was found in \cite{Ashok:2002bi} that the effective superpotential of the ${\cal N}=1$ SYM gauge theory is fully captured by the contributions from $S^2$ and $\bR \bP^2$, so there is no contribution from the world-sheet with two crosscaps (Klein bottle); furthermore, the contribution to the free energy coming from $\bR \bP^2$ is given by a derivative of the contribution from $S^2$,
$$
{\cal F}_1=\pm \frac{g_s}{4} \frac{\partial {\cal F}_0}{\partial S_0}
$$
with $S_0$ (half) the 't Hooft coupling. Notice however that in this example the properties are only established for world-sheets without any handles or boundaries, while our arguments work for world-sheets with a single boundary and an arbitrary number of handles.  A second example comes from the large N expansion of Chern-Simons theory on 3-manifolds. It was observed in \cite{Sinha:2000ap} that the $1/N$ expansion of the free energy of Chern-Simons on $S^3$ with gauge groups $SO(N),Sp(N)$ involves unoriented world-sheets with one cross-cap, but again world-sheets with two cross-caps are absent in this expansion. Moreover, the large N expansion of Chern-Simons with $G=SO(N),Sp(N)$, via its connection with knot theory, displays non-trivial relations for the invariants of $U(N)$ and $SO(N),Sp(N)$  links \cite{Marino:2009mw}.

While it is interesting that the two features we have uncovered in the $1/N$ expansion of $\vev{W(g)}_{SO(2N)}$ and $\vev{W(g)}_{Sp(N)}$ have superficially similar incarnations in other gauge theories with gauge groups $SO(N), Sp(N)$, we don't expect these two features to be generic for other observables of ${\cal N}=4$ SYM with $G=SO(N),Sp(N)$. For instance, in the case we have studied, the absence of contributions coming from world-sheets with two crosscaps is a consequence of the exact relation (\ref{exactsum}), but this relation appears to be quite specific of vevs of Wilson loops in the respective fundamental representations, and we don't know of similar relations for vevs of Wilson loops in other representations. Not surprisingly, in Chern-Simons theory with $G=SO(N),Sp(N)$, vevs of Wilson loops in higher representations do get contributions from world-sheets with two crosscaps \cite{Bouchard:2004iu}.

Turning now to string theory, reproducing the actual $1/N$ expansion of $\vev{W(g)}_{SO(2N)}$ or $\vev{W(g)}_{Sp(N)}$ from world-sheet computations is as out of reach as for $\vev{W(g)}_{U(2N)}$. On the other hand, granting the AdS/CFT duality for any value of $g_s$ and $\alpha'/L^2$, our results are also exact results in string theory, even beyond the perturbative regime. It is tantalizing to suspect that the results we have found - {\it e.g.} the absence of contributions from world-sheets with two crosscaps and any number of handles - are in the string theory language consequences of some symmetry enjoyed by the particular quantities we are considering. Identifying this symmetry and the stringy argument beyond the relations we have found appears to be a more promising  and illuminating task than attempting to reproduce them by carrying out the explicit world-sheet computations.

Everything we have said so far follows from the exact results we have computed, and the exact relations among them. We didn't even have to carry out the explicit $1/N$ expansion of the exact results to arrive at these conclusions. Nevertheless, it is still worth to obtain this $1/N$ expansion explicitly, and this task can be accomplished with very little effort, by combining the exact relation (\ref{exactdiff}) with the results in \cite{Drukker:2000rr}. Drukker and Gross  \cite{Drukker:2000rr}  obtained the following $1/N$ expansion of $\vev{W}_{U(N)}$, that we write for $U(2N)$, 
$$
\vev{W}_{U(2N)}=\frac{2}{\sqrt{2\lambda}}I_1(\sqrt{2\lambda})+\sum_{k=1}^\infty\frac{1}{N^{2k}}\sum_{i=0}^{k-1}X_k^i\left(\frac{\lambda}{2}\right)^{\frac{3k-i-1}{2}}I_{3k-i-1}(\sqrt{2\lambda})
$$
where $I_\alpha(x)$ are modified Bessel functions of the first kind, and $X_k^i$ are coefficients satisfying the recursion relation
\begin{equation}
4(3k-i)X_k^i=X_{k-1}^i+(3k-i-2)X_{k-1}^{i-1}
\label{recur}
\end{equation}
with initial values $X_1^0=1/12$ and $X_k^k=0$. A trivial integration then yields 
$$
\vev{W}_{\substack{SO(2N)\\ Sp(N)}}=\vev{W}_{U(2N)}\mp \frac{1}{4N}\left[\left(I_0(\sqrt{2\lambda})-1\right)+\sum_{k=1}^\infty\frac{1}{N^{2k}}\sum_{i=0}^{k-1}X_k^i\left(\frac{\lambda}{2}\right)^{\frac{3k-i}{2}}I_{3k-i}(\sqrt{2\lambda})\right]
$$
This result is valid for any $\lambda$. We can then use it to obtain a large $\lambda$ expansion at every order in $1/N$
$$
\vev{W}_{SO(2N)}-\vev{W}_{Sp(N)}=
\sum_k \frac{1}{(2N)^{2k+1}} \frac{e^{\sqrt{2\lambda}} (2\lambda)^{\frac{6k-1}{4}}}{96^k k! \sqrt{2\pi}}\left(1-\frac{36k^2+144k-5}{40\sqrt{2\lambda}}+\dots\right)
$$
Perhaps the most important feature of this result is that the exponent $(6k-3)/4$ obtained in \cite{Drukker:2000rr} is now replaced by $(6k-1)/4$. 

\section{Acknowledgements} 
We would like to thank Ofer Aharony, Pablo G. C\'amara, Nadav Drukker and Marcos Mari\~no for useful comments and correspondence. The research of BF is supported by MEC FPA2010-20807-C02-02, CPAN CSD2007-00042, within the Consolider-Ingenio2010 program, and AGAUR 2009SGR00168. BG is supported by an ICC scholarship and by MEC FPA2010-20807-C02-02, and GT by an FI scholarship by the Generalitat the Catalunya.

\appendix
\section{Classical simple Lie algebras}
In this appendix we collect some very basic facts about classical simple Lie algebras that we use in the main text. A Lie algebra of rank $r$ has $r$ simple roots. For each simple root in the Lie algebra there is a fundamental weight, which is the highest weight of a fundamental representation. A simple Lie algebra has then $r$ fundamental representations. In Physics, the name ``fundamental representation" if often reserved for the fundamental representation with highest weight $w_1$.
\bigskip

\underline{$\mathfrak{su(n)}$} The Lie algebra $\mathfrak{su(n)}$ has rank $r=n-1$. We introduce the basis $e_i, i=1,\dots,n$. The positive roots and the simple roots are
\begin{align*}
& R_+  =  \{e_i-e_j, i<j \} \\
& \Pi  =  \left\{\alpha_1=e_1-e_2,\dots,\alpha_{n-1}=e_{n-1}-e_n \right\}
\end{align*}
The $n-1$ fundamental weights of ${\mathfrak{su(n)}}$ are
\begin{equation}
w_k=e_1+e_2+\dots+e_k-\frac{k}{n}\left(e_1+\dots+e_n\right),\hspace{1cm}k=1,\dots,n-1
\label{fundwei}
\end{equation}
Applying the Weyl dimension formula, the dimensions of the associated fundamental representations are ${n \choose k}$, so these are the antisymmetric representations.
\bigskip

\underline{$\mathfrak{so(2n+1)}$}. The Lie algebra $\mathfrak{so(2n+1)}$ has rank $r=n$. We introduce the basis $e_i, i=1,\dots,n$. The positive roots and  the simple roots are
\begin{align*}
& R_+=\{e_i\pm e_j \; (i<j) ,e_i\} \\ 
& \Pi=\{\alpha_1=e_1-e_2,\dots,\alpha_{n-1}=e_{n-1}-e_n,\alpha_n=e_n\}
\end{align*}
The fundamental weights are
\begin{flalign*}
& w_1=e_1,\dots,w_{n-2}=e_1+\dots+e_{n-2},w_{n-1}=e_1+\dots+e_{n-1}, \\
& w_n=\frac{1}{2}\left(e_1+\dots+e_n\right)
\end{flalign*}
The first $n-1$ representations have dimensions ${2n+1 \choose k}$. The last one is a spinor representation of dimension $2^n$.
\bigskip

\underline{$\mathfrak{sp(n)}$}. The Lie algebra $\mathfrak{sp(n)}$ has rank $r=n$. We introduce the basis $e_i, i=1,\dots,n$. The positive roots and the simple roots are
\begin{align*}
&R_+=\{e_i\pm e_j \; , \; i<j  \; ; 2 e_i\}
&\Pi=\{\alpha_1=e_1-e_2,\dots,\alpha_{n-1}=e_{n-1}-e_n,\alpha_n=2e_n\}
\end{align*}
The corresponding fundamental weights are
$$
w_1=e_1,w_2=e_1+e_2, \dots, w_n=e_1+\dots+e_n
$$
There are no spinor representations for $\mathfrak{sp(n)}$.
\bigskip

\underline{$\mathfrak{so(2n)}$}. The Lie algebra $\mathfrak{so(2n)}$ has rank $r=n$. We introduce the basis $e_i, i=1,\dots,n$. The positive roots and the simple roots are
\begin{align*}
& R_+=\{e_i\pm e_j \;, \; i<j \;\} \\
& \Pi=\{\alpha_1=e_1-e_2,\dots,\alpha_{n-1}=e_{n-1}-e_n,\alpha_n=e_{n-1}+e_n\}
\end{align*}
The corresponding fundamental weights are
\begin{flalign*}
& w_1=e_1\; , w_2=e_1+e_2 \; , \dots, \; w_{n-2}=e_1+\dots+e_{n-2},  \\
&  w_{n-1}=\frac{1}{2}\left(e_1+\dots+e_{n-1}-e_n\right),
w_n=\frac{1}{2}\left(e_1+\dots+e_{n-1}+e_n\right) 
\end{flalign*}
The first $n-2$ fundamental representations have dimensions ${2n\choose k}$. The last two fundamental weights correspond to spinor representations, both with dimension $2^{n-1}$. 

\section{1$/$N expansion of $\vev{W(g)}_{SO(2N)}$ and $\vev{W(g)}_{Sp(N)}$}
In this appendix we will derive the $1/N$ expansion of $\vev{W(g)}_{SO(2N)}$ and $\vev{W(g)}_{Sp(N)}$ without making use of the exact relations among them found in the main text. We will eventually find out that the expansions involve certain coefficients that satisfy the same recursion relation as the ones that appear in $\vev{W(g)}_{U(2N)}$, eq. (\ref{recur}). 

To expand $\vev{W(g)}_{SO(2N)}$ given in eq. (\ref{exactso}) in $1/N$, we will first rewrite 
$$
\sum_{k=0}^{N-1}L_{2k}(-g) =\sum _{k=0}^{2N-2} d_k \frac{g^k}{k!}
$$
with 
$$
d_k \equiv \sum_{i=0}^{N-1}{2i \choose k}
$$
These coefficients satisfy the recursion relation
$$
d_k+2d_{k+1}={2N \choose k+2}
$$
and with $d_0=N$ we can now write
$$
\vev{W(g)}_{SO(2N)}=\frac{1}{N}\sum_{n=0}^\infty \left(\frac{\lambda}{2}\right)^n \frac{1}{n! (n+1)!} D(n,N)
$$
with 
$$D(n,N)\equiv 2\frac{n!(n+1)!}{(2N)^{n+1}}\sum_{k=0}^n\frac{d_k}{2^{n-k}(n-k)! k!}
$$
$D(n,N)$ is a polynomial in $1/N$ of degree $n$. Expanding in $1/N$,
$$
D(n,N)=1-\frac{n+1}{2}\frac{1}{2N}+\frac{(n+1)n(n-1)}{12}\frac{1}{\left(2N\right)^2}+\dots
$$
So
$$
\vev{W(g)}_{SO(2N)}=\sqrt{\frac{2}{\lambda}}I_1(\sqrt{2\lambda})-\frac{1}{4N}\left(I_0(\sqrt{2\lambda})-1\right)+\dots
$$

To expand $\vev{W(g)}_{Sp(N)}$ given in eq. (\ref{exactsp}) in $1/N$, we will first rewrite 
$$
\sum_{k=0}^{N-1}L_{2k+1}(-g) =\sum _{k=0}^{2N-1} c_k \frac{g^k}{k!}
$$
with 
$$
c_k \equiv \sum_{i=0}^{N-1}{2i+1 \choose k}
$$
These coefficients satisfy the recursion relation
$$
c_k+2c_{k+1}={2N+1 \choose k+2}
$$
and with $c_0=N$ we can now write
$$
\vev{W(g)}_{Sp(N)}=\sum_{n=0}^\infty \left(\frac{\lambda}{2}\right)^n \frac{1}{n! (n+1)!} C(n,N)
$$
with 
$$C(n,N)\equiv 2\frac{n!(n+1)!}{(2N)^{n+1}}\sum_{k=0}^n\frac{c_k}{2^{n-k}(n-k)! k!}
$$
$C(n,N)$ is a polynomial in $1/N$ of degree $n$. Expanding in $1/N$,
$$
C(n,N)=1+\frac{n+1}{2}\frac{1}{2N}+\frac{(n+1)n(n-1)}{12}\frac{1}{\left(2N\right)^2}+\dots
$$
So
$$
\vev{W(g)}_{Sp(N)}=\sqrt{\frac{2}{\lambda}}I_1(\sqrt{2\lambda})+\frac{1}{4N}\left(I_0(\sqrt{2\lambda})-1\right)+\dots
$$
We know from general arguments that the odd powers in $1/N$ of $C(n,N)$ and $D(n,N)$ differ by a sign. Now we want to argue that the even powers are the same, so as polynomials in $1/N$ we have $D(n,-N)=C(n,N)$. Define
$$
\Delta(n,N)\equiv C(n,N)-D(n,N)=2\frac{n!(n+1)!}{(2N)^{(n+1)}}\sum _{k=1}^n\frac{d_{k-1}}{2^{n-k}(n-k)! k!}
$$
If we prove that $\Delta(n,N)$ is a polynomial in $1/N$ with only odd powers, it will follow that even powers of $C$ and $D$ coincide. The coefficients $\Delta$ satisfy the recursion relation
$$
\Delta(n+1,N)=\frac{n+2}{n+1}\Delta(n,N)+\frac{(n-1)(n+2)}{16N^2}\Delta(n-1,N)
$$
Together with $\Delta(0,N)=0, \Delta(1,N)=1/N$ this proves that indeed $\Delta(n,N)$ are odd in $1/N$, and indeed even powers of $C$ and $D$ coincide.

To carry out the expansion of $\Delta(n,N)$ we follow closely Appendix B of \cite{Drukker:2000rr}. We define
$$
\Delta(n,N)=\sum_k \frac{p_k(n)}{(2N)^{2k+1}}
$$
where $p_k(n)$ are polynomials in $n$ of degree $3k+1$. We rewrite them as linear combinations of polynomials $(n+1)!/(n-3k+i)!$ with coefficients $Y_k^i$,
$$
p_k(n)=\sum_{i=0}^{k-1}\frac{(n+1)!}{(n-3k+i)!} Y_k^i
$$
Using the recursion relation for $\Delta(n,N)$ we derive the relation
$$
4(3k-i)Y_k^i=Y_{k-1}^i+(3k-i-2)Y_{k-1}^{i-1}
$$
which is the same recursion relation found in  \cite{Drukker:2000rr} for the cofficients $X_k^i$, eq. (\ref{recur}). The initial values can also be seen to coincide, proving that the unoriented term are related to the oriented ones.

\end{document}